\documentclass{aa}
\usepackage{amsmath}
\usepackage{xspace}
\usepackage{graphicx}
\usepackage{amssymb}
\usepackage{natbib}

\usepackage{xspace}

\newcommand{\micron}{\unskip~\ensuremath{\mu}m\xspace}
\newcommand{\ten}[1]{10\ensuremath{^{#1}}\xspace}
\newcommand{\Msun}{M\ensuremath{_{\odot}}\xspace}
\newcommand{\Rsun}{R\ensuremath{_{\odot}}\xspace}
\newcommand{\Lsun}{L\ensuremath{_{\odot}}\xspace}

\newcommand{\coo}{\ensuremath{\rm{CO}_{2}}\xspace}

\def\cdrev#1{#1}
\def\cdrevtwo#1{{\bf #1}}
\def\kel{\hbox{K}}
\def\fullstop{\,.}
\def\comma{\,,}
\def\AU{\hbox{AU}}

\def\rin{{R_{\mathrm{in}}}}

\def\dust{\hbox{\scriptsize d}}
\def\disk{\rm disk}
\def\Tcenter{\ensuremath{T_{\mathrm{c}}}}
\def\hnull{h_0}
\def\hs{H}
\def\hp{h_p}
\def\settl{\rm set}
\def\kepler{\rm k}
\def\drift{\rm dr}

\begin{document}
\title{The dust disk of HR4049}
\subtitle{Another brick in the Wall}
\author{C.~Dominik\inst{1}, C.P.~Dullemond\inst{2}, J.~Cami\inst{1,3} and H.~van Winckel\inst{4}}
\authorrunning{Dominik et al.}
\offprints{C. Dominik}

\institute{Sterrenkundig Instituut `Anton Pannekoek', Kruislaan 403,
  NL-1098 SJ Amsterdam, The Netherlands;\\ e--mail: dominik@science.uva.nl 
\and
Max Planck Institut f\"ur Astrophysik, Karl
Schwarzschild Strasse 1, D--85748 Garching, Germany;\\ e--mail:
dullemon@mpa-garching.mpg.de
\and
SRON-Groningen, PO Box 800, 9700 AV Groningen, The Netherlands;\\
e--mail:  cami@astro.rug.nl
\and
Instituut voor Sterrenkunde, Katholieke Universiteit Leuven,
Celestijnenlaan 200 B, B-3001 Heverlee, Belgium;\\ e--mail: Hans.VanWinckel@ster.kuleuven.ac.be}
\date{\today}

\abstract{ We present the Spectral Energy Distribution of HR 4049
  based on literature data and new continuum measurements at 850
  \micron. The SED shows variable absorption in the UV, and a large IR
  excess, both caused by circumstellar dust.  The shape of the IR
  excess from 1 \micron all the way down to 850 \micron can be nearly
  perfectly fitted with a single blackbody function at $T \approx
  1150$ K or alternatively with a sum of blackbodies in a narrow
  temperature range. The energy emitted in this IR continuum radiation
  is about one-third of the stellar luminosity. We show that this
  blackbody radiation must be due to the presence of a circumbinary
  disk with a large height. This disk must also be gas-rich, in
  agreement with the observations of molecular bands in the ISO-SWS
  spectrum.  We present two possible scenario's for explaining the
  shape and the intensity of the IR excess. The first scenario
  involves large grains ($a \ge $ 1 mm) that each radiate like a
  blackbody. The second scenario argues that the blackbody radiation
  is due to a very optically thick circumbinary disk. We investigate
  if such a disk would indeed produce blackbody radiation by
  presenting results from radiative transfer calculations.  We further
  quantify the properties of such a disk and its stability in the
  framework of (hydro)dynamics, grain settling, radiation pressure and
  grain drift.  The virtues and shortcomings of both models for the
  origin of the IR blackbody are discussed by contrasting them with
  other observations and assessing them in the framework of (binary)
  \mbox{(post-)AGB} evolution.
}

\maketitle

\section{Introduction}

When low- to intermediate mass stars leave the Asymptotic Giant Branch
(AGB), they spend a relatively short time (\ten{3} to \ten{4} years)
on their way to become a planetary nebula; this stage of stellar
evolution is the post-AGB phase. During this phase, the stellar
interior is gradually more exposed, so that the effective temperatures
increase from typical AGB values ($\sim$ 3000 K) to those commonly
found in planetary nebulae ($\sim$ 30 000 K). At the same time, the
luminosity of these stars remains more or less constant and is
generally in the range \ten{3} - $5\times$ \ten{4} \Lsun. As the AGB
phase is characterized by substantial mass-loss, it is expected that
post-AGB stars are \cdrevtwo{surrounded by} circumstellar matter. It is this
material that might later on become visible as a
planetary nebula. \\
HR4049 is considered to be the prototype of a group of post-AGB stars
sharing similar - but rather peculiar - characteristics
\citep[see][]{hans:post-AGB-binaries}. They are all members of a
(close) binary system, often with highly eccentric orbits. Their
photospheres show a high degree of metal-depletion, while volatile
elements as CNO, S and Zn show solar-like abundances. Most of them
exhibit a deficiency in the UV and optical, while they do have a -
sometimes huge - IR excess, pointing to the presence of circumstellar
dust which thermally re-emits the radiation it absorbs in the UV. The
key to explaining all of these peculiarities is probably the binary
nature of these objects which allows the formation of a massive,
stable and long-lived circumbinary disk. The formation of dust in
these disks followed by re-accretion of the depleted gas onto the
stellar photosphere explains the photospheric abundance patterns
\cdrev{\citep{rens:selective-depletion}}.
Moreover, the interplay of a massive circumbinary disk with the
central binary might be important for the dynamics of the binary
system \citep{1997A&A...321L..29M,1988A&A...202..177I}. The lead role
of the dust disks
in these systems can therefore hardly be overestimated. \\
HR4049 was first suggested to be an object in the transition phase
from the AGB to the Planetary Nebula stage by \citet{lamers:HR4049}
who discovered both a large IR excess and severe UV deficiency
indicating the presence of circumstellar dust. From a variety of
observational results, it is clear that the properties of this
circumstellar dust are unusual. \\
\citet{waelkens:HR4049-variability} showed that photometric variations
are correlated with radial velocity variations :  the star is faintest
and reddest when the star is at inferior conjunction. The most plausible explanation for this phenomenon is that
the dust which is responsible for the extinction is not distributed in
a spherical shell, but concentrated in a disk \cdrev{surrounding} the entire
binary system -- a circumbinary disk -- observed at a rather high
inclination : in such a geometry the amount of dust in the line of
sight is largest at inferior conjunction which explains both the
photometry and color variations. \\
The amplitude of the circumstellar extinction is wavelength-dependent,
and scales linearly with $\lambda^{-1}$
\citep{waelkens:HR4049-variability,rens:HR4049-dust,Buss:HR4049-extinction}. 
The extinction remains linear in the UV and shows no sign of the 2175
\AA\ bump \citep{Buss:HR4049-extinction}. A linear spectral extinction
can be understood in terms of Mie scattering when the size of the dust
grains is small compared to the wavelength. This yields an upper limit
of 300 \AA\ for the size of the dust grains in the line-of-sight
towards HR4049. 
\citet{joshi:hr4049-polarization} and
\citet{johnson:hr4049-polarization} presented spectropolarimetry in
the UV and the optical. The data show a high degree of
polarization in the UV, with a change of 90$\degr$ in the position
angle around 2000 \AA - the signature of scattering material in a
bipolar distribution. The polarization in the optical seems to vary with
orbital phase \citep{johnson:hr4049-polarization}, suggesting that the
polarization in the optical is due to scattering in the circumbinary
disk. In the UV, the polarization is then caused by scattering in the
bipolar lobes, which should contain a population of small grains ($a<$
0.05 \micron). \\
\citet{rens:HR4049-dust} studied the properties of the circumstellar
dust around \object{HR4049}. They detected emission features
attributed to Polycyclic Aromatic Hydrocarbons (PAHs); the ISO/SWS
spectrum of this object \citep{douwe:circumstellar-PAHs} shows these
features in great detail, suggestive of a C-rich dust composition.
This is further corroborated by \citet{Geballe:HR4049-IRfeatures} who
studied the profiles of the 3.43 and 3.53 \micron\ emission features
recently identified with vibrational modes of hydrogen-terminated
crystalline facets of diamond \citep{Guillois:nanodiamonds}.\\
In an accompanying paper (Paper I hereafter, in preparation) we
analyze the molecular bands found in the ISO/SWS spectrum. Although
the dust features present in the spectrum can all be attributed to
C-rich components, the molecular species are typical for the
composition found around an O-rich AGB star.  The dust surrounding
HR 4049 therefore seems to be C-rich, while the gas is more O-rich in
nature. \\
Another intriguing property of the ISO/SWS spectrum is that the
entire IR continuum can be matched nearly perfectly with a blackbody
at a single temperature of 1200 K. An extrapolation of this blackbody
also matches the 60 \micron IRAS point. \\  
In this paper we (re-)investigate the circumstellar environment
of HR 4049. In Sect.~\ref{sec-the-sed} we present the Spectral Energy
Distribution (SED) of HR 4049 from the UV to submm wavelengths,
showing that the dust continuum remains a blackbody to the longest
wavelengths observed so far. In Sect.~\ref{sec:constraints} we show
that the SED can be used to derive strong constraints on the
circumstellar dust distribution. We conclude that there are only two
possible dust distributions - both having a disk geometry - that can
meet these observational constraints; these are described in
Sect.~\ref{sec:optically_thin} and \ref{sec:optically_thick}. The
properties of the circumstellar dust disk within the framework of
these two possible scenarios are addressed in
Sect.~\ref{sec:disk-properties}. In Sect.~\ref{sec:discussion} we
compare the two possibilities in the framework of disk dynamics and
stellar evolution, and assess the plausibility of both models in the
framework of other observations presented in the literature. 

\section{The Spectral Energy Distribution of HR4049}\label{sec-the-sed}

\begin{figure*}
\resizebox{\hsize}{!}{\includegraphics[clip=]{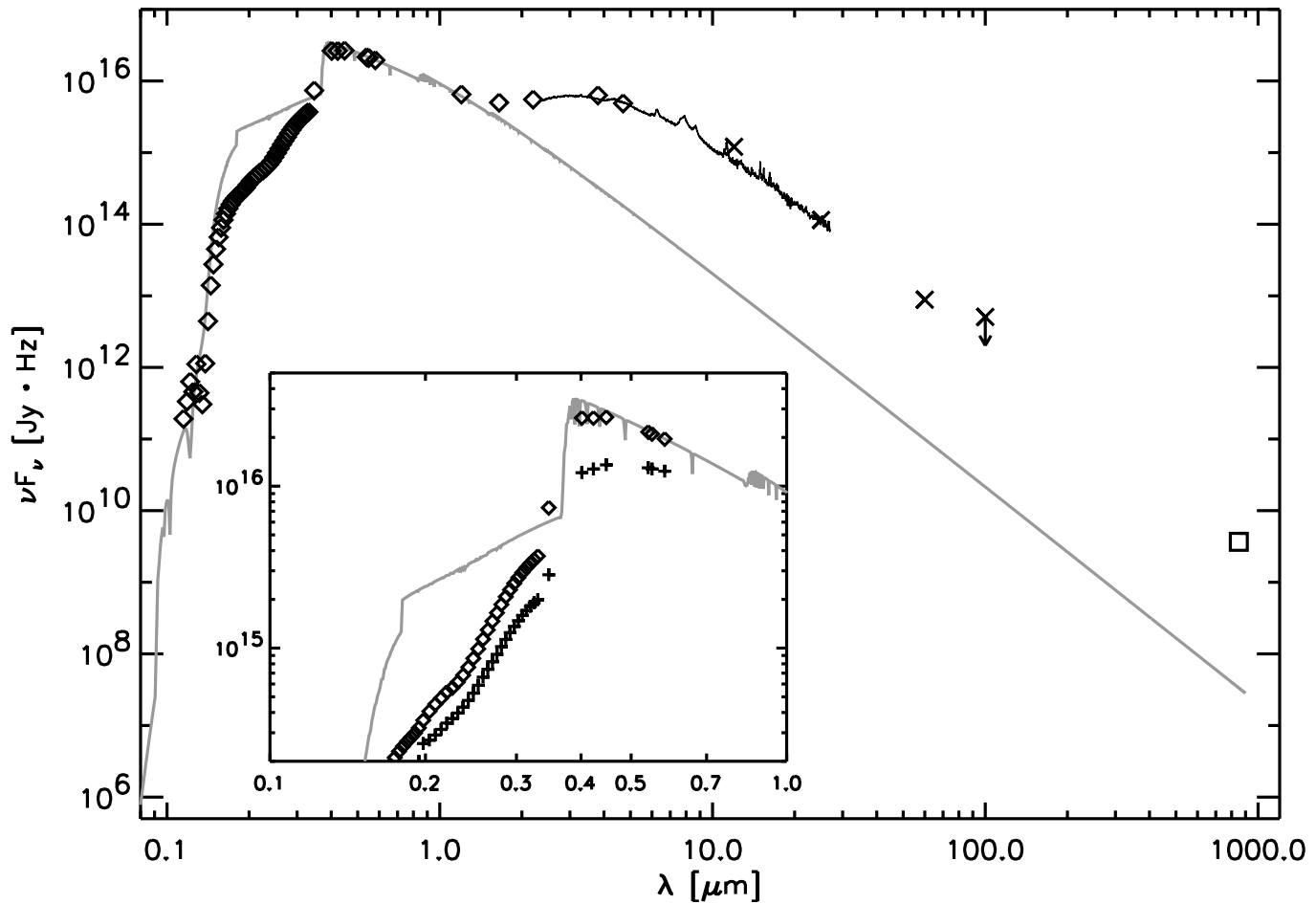}}
\caption{\label{fig:sed}The SED of HR 4049 at photometric maximum. 
  The diamonds show IUE measurements, optical and near-IR photometry;
  these measurements are de-reddened using $E(B-V)=0.1$. 
  IRAS observations at 12, 25, 60 and 100 \micron are shown as
  crosses. Note that the 100 \micron point is only an upper limit. The
  continuum measurement at 850 \micron based on the new SCUBA
  observations is represented as the square. 
  The ISO-SWS spectrum between 2.3 and 27 \micron is
  plotted as the solid black line. 
  The thick grey line shows a Kurucz model for the star with
  parameters $T_{\mathrm{eff}}=7500\kel$, $\log g = 1$ and
  $[Fe/H]=-4.5$. This Kurucz model is scaled to match
  the SED at photometric maximum at 580 nm. The inset shows the SED at
  photometric 
  maximum (diamonds) and photometric minimum (+) at UV and
  optical wavelengths, revealing (variable) extinction from
  circumstellar dust.} 
\end{figure*}
To compose the Spectral Energy Distribution (SED) of HR 4049, we used
UV, optical and near-IR observations found in the literature, data
obtained with the Short Wavelength Spectrometer
\citep[SWS,][]{deGraauw:SWS} on board the Infrared Space
Observatory \citep[ISO,][]{kessler:ISO} and new data obtained with the
James Clerk Maxwell Telescope (JCMT). \\
The UV part of the SED is based on the many measurements of HR 4049 by
the International Ultraviolet Explorer (IUE) \citep[see
e.g.][]{Monier:HR4049-UV}. Ground-based optical and near-IR photometry
observations were acquired over a long period; the observations we use
in this paper are basically the same as presented in
e.g. \citet{hans:post-AGB-evolution}. In the IR, we use the ISO/SWS
spectrum from 2.3 to 30 \micron (see Paper I) and IRAS photometry
points at 12, 25, 60 and 100 \micron. Note however that the 100
\micron point is an upper limit; this observation is probably
contaminated by interstellar cirrus clouds. \cdrev{The 60
  \micron point is not contaminated because the emission at this
  wavelength is clearly a point source \citep{rens:HR4049-dust}.} \\ 
In this paper we report upon very recent observations carried out with
the JCMT using the Submillimetre
Common-User Bolometer Array (SCUBA) to obtain a continuum measurement
at 850 \micron. HR 4049 was observed 3 times (31 Jan 2001, 9 Oct
2001 and 23 Nov 2001). The data were reduced using Uranus as a
calibrator; the atmospheric opacity at 850 \micron is fitted with a
straight line at epochs between the observations of photometric
calibrators. The total integration time was 2700 seconds and yields a
flux of 10.4 $\pm$ 2.2 mJy. We included this point in the SED. \\
As HR 4049 shows considerable photometric variations from the UV to
the near-IR \citep[e.g.][]{waelkens:HR4049-variability}, we composed a
SED for both photometric maximum and for photometric minimum. \\
\cdrevtwo{The
  best estimate for the distance to HR\,4049 is given by the Hipparcos
parallax of $0.150 \pm 0.064$ mas which corresponds to a distance range
between 467 and 1162 pc.}
To study the properties of the circumstellar dust, we first need to
correct the SED for interstellar extinction. This correction is
somewhat uncertain, as the extinction towards HR 4049 contains both a
(variable) circumstellar and an interstellar component.  As the color
variations are in phase with the photometric and radial velocity
variations, the circumstellar reddening is lowest at photometric
maximum. To minimize the effects of circumstellar extinction, we
therefore used the SED at photometric maximum to estimate the
interstellar reddening. \\
We estimated the interstellar component of the $E(B-V)$ by using the
average interstellar extinction law given by
\citet{Savage-Mathis:IS-extinction} to de-redden the observed SED at
photometric maximum and subsequently comparing this de-reddened SED
with a Kurucz model. Several authors derived stellar parameters for HR
4049 using different methods \citep[see][for an
overview]{bakker:hr4049-variability}; we used these parameters and
compared the de-reddened SED therefore with a Kurucz model with
$T_{\mathrm{eff}}=7500\kel$, $\log g=1$ and $[Fe/H]=-4.5$. We found
that adopting an interstellar reddening of $E_{\rm IS}$=0.1 yields the
best fit in the optical between the SED at photometric maximum and the
Kurucz model. This value for the interstellar reddening toward HR 4049
agrees well with the reddening determined for nearby stars which was
found to be  $0.09 < E_{\rm IS}(B-V) < 0.17$
\citep{Buss:HR4049-extinction}. We adopted the same value for the
$E_{\rm IS}(B-V)$ at photometric minimum. 
Fig.~\ref{fig:sed} shows the resulting de-reddened SEDs at photometric
maximum and minimum and the Kurucz model.  \cdrevtwo{Note that the
  plot shows a clear UV of the dereddened spectrum compared to the
  Kurucz model with appropriate metallicity, a clear sign of anomalous
  circumstellar extinction which is also variable with orbital phase
  \citep{1995Ap&SS.224..357W,Monier:HR4049-UV}.}
In Fig.~\ref{fig:subfit} we have subtracted the
stellar model from the
SED in order to show only the IR excess. \\
\section{Constraints on the dust distribution}
\label{sec:constraints}
\subsection{Energy considerations}
\label{sec:energy}

The SED shows a UV deficiency and an IR excess, both caused by
circumstellar dust. The circumstellar dust absorbs the stellar
radiation in the UV producing the UV deficiency.  The IR excess is
caused by grains which are heated by stellar radiation and re-emit
this energy thermally.  It is important to realize (and often
overlooked) that the IR excess is produced by all emitting dust grains
in the beam, while the UV deficit is only due to absorption by dust
grains {\em in the line-of-sight}. The source of the UV deficit and
the IR excess are therefore not necessarily the same dust
particles. \\
To get a feeling for the amount of absorption in the UV and the IR
emission of the dust, we calculate the amount of energy absorbed
in the UV and emitted in the IR by numerically integrating (parts of)
the SEDs and the Kurucz model and express these energies in units of
the stellar flux. All integrations were done using a five-point
Newton-Cotes integration formula.  \\
We first calculated the stellar flux $F_{\star}$ by integrating the
Kurucz model between 145 nm and 850 \micron. The energy radiated at
shorter and longer wavelengths is only a negligible fraction ($\la$
\ten{-4}) of the total stellar flux.  To calculate the amount of
energy absorbed, we should integrate the SEDs without the IR excess at
both photometric maximum and minimum in order to find the fluxes
$F^{\rm max}$ and $F^{\rm min}$, respectively.  \cdrev{The IR excess
dominates at wavelengths longer than 1.2\micron; the
stellar flux at these wavelengths however still contributes about 10 \% of
the total stellar luminosity.  Fortunately the extinction becomes small at
those \cdrevtwo{wavelengths}. We therefore assumed that the intrinsic
SED at these 
wavelengths is equal to the Kurucz model.}  We then find
\begin{equation}
\label{eq:max_abs}
1 - \frac{F^{\rm max}}{F_{\star}} = 0.04
\end{equation}
\begin{equation}
\label{eq:min_abs}
1 - \frac{F^{\rm min}}{F_{\star}} = 0.38
\end{equation}
where the superscripts refer to photometric maximum and minimum
respectively. At photometric maximum, the dust in the line-of-sight
thus absorbs 4 \% of the stellar radiation while at photometric
minimum this amounts to 38 \%. 
We can also use the SED to get an estimate of the optical depth : 
\begin{equation}
\label{eq:tau_from_SED}
\tau = - \ln \frac{I_{\rm obs}}{I_{\rm Kurucz}}
\end{equation}
If the optical depth were grey, we can use Eqs.~(\ref{eq:max_abs}) and
(\ref{eq:min_abs}) to find that $\tau_{\rm max} \approx 0.04$ and
$\tau_{\rm min} \approx$ 0.48. However, the extinction in the UV is
not grey \citep{Buss:HR4049-extinction} and therefore the optical
depth in the UV will be larger than these grey values. 
We used Eq.~\ref{eq:tau_from_SED} to get an idea about the actual
(wavelength-dependent) optical depth and found that the optical depth
reaches maximum values of $\tau \sim$ 1.5 - 2 between 180 and 200 nm at
photometric maximum, and even larger values ($\tau \sim$ 2 - 2.5) at
photometric minimum. \\  
Integrating over the IR excess yields
\begin{equation}
\label{eq:IR_excess}
\frac{F_{\mathrm{IR}}}{F_{\star}} \approx 0.32 \approx \frac{1}{3}
\end{equation}
so the energy radiated by the dust in the IR is about one third of the
energy radiated by the star at all wavelengths!  It therefore
seems that the UV deficiency at minimum light and the IR
excess both amount to about one third of the stellar flux.
Looking only at the observations at minimum light, one might
therefore be tempted to conclude that the dust is distributed
spherically around the star and that the line-of-sight we probe with
the UV absorption is representative.  However, the large variation
of the UV extinction with phase in fact tells us that the similar
numbers are purely coincidental.  We will get back to this in the
discussion.
\begin{figure}
\resizebox{\hsize}{!}{\includegraphics{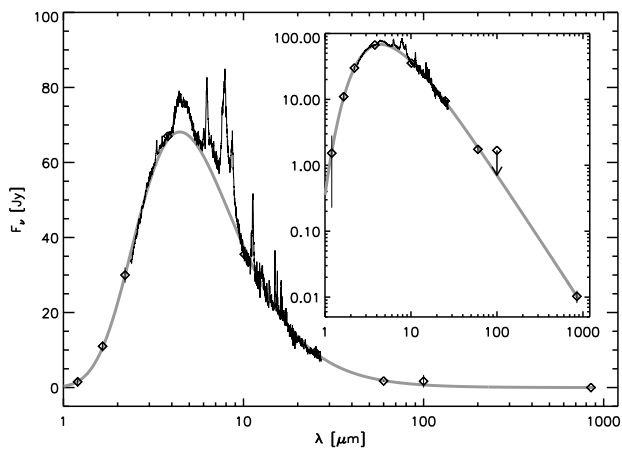}}
\caption{\label{fig:subfit}The IR-excess of HR 4049. The Kurucz model
  is subtracted from the SED shown in Fig.~\ref{fig:sed}. Errors on
  the near-IR photometry points, the IRAS measurements and the SCUBA
  point are indicated, but are generally smaller than the symbol
  size. Note that the ``photometric'' points at 3.8 \micron and 10
  \micron were derived from the ISO spectrum (see text for more
  details). The prominent features in the ISO-SWS spectrum are mainly
  due to PAHs, nano-diamonds, and \coo (see Paper I). 
  The thick grey line shows a single blackbody at a temperature of 1150 K.
  The inset shows that this single temperature blackbody provides a
  perfect fit to the entire continuum, including the 60\micron
  and 850\micron measurements.}
\end{figure}

\subsection{Dust Temperatures}
\label{sec:temperatures}
In Fig.~\ref{fig:subfit} we have subtracted the stellar model from
the SED in order to show only the IR dust emission.  It shows a shape
remarkably close to a blackbody over almost 3 orders of magnitude in
wavelength, from 1.2 to 850 \micron. As we showed in Paper I, a Planck
function at a single temperature of $T \simeq$ 1200 provides an
excellent fit to the IR continuum up to 60 \micron. In this Section, we
will further quantify and constrain this dust temperature. \\
The first question to answer is if a single temperature blackbody can
really provide a good fit to the continuum in the statistical sense. 
We therefore decided to perform a $\chi^2$ minimization. As the ISO-SWS
spectrum shows a wealth of features due PAHs, nano-diamonds and
molecular bands (see Paper I), we preferred not to use the spectrum
for this purposes and only restrict ourselves to the near-IR
photometry observations, the IRAS measurements and the SCUBA
point. However, when comparing the L and M band observations (3.8 and
4.7 \micron) with the ISO spectrum, it is seen that the fluxes of the
photometric points are slightly higher than the ISO-SWS continuum;
the photometric points are likely influenced by the strong emission
from the \coo stretching band which is centered around 4.2
\micron. Also the 12 \micron IRAS point is not representative for the
continuum (see Fig.~\ref{fig:sed}); in this case, the culprit is
probably the strong 11.6 PAH feature. Given these considerations, we
decided not to use these photometric points. Instead, we used the
ISO-SWS spectrum to obtain ``photometric'' continuum points around
these wavelengths. We replaced the L band photometric point with the
mean flux in the ISO-SWS spectrum between 3.7 and 3.9 \micron; this
range seems to be free of spectral features. The 12 \micron IRAS
measurement was replaced by the mean flux between 10 and 10.2
\micron. It is somewhat uncertain whether this is really a good
continuum point, as the PAH features might still contribute to the
flux at these wavelengths. It is however a much better continuum point
than the IRAS 12 \micron one. \\
We next scaled a single temperature blackbody to the K-band
flux (2.2 \micron) and calculated $\chi^2_{\nu}$ (reduced
$\chi^2$). We found that a single temperature blackbody can indeed
accurately reproduce the continuum ($\chi^2_{\nu} < 1$) provided the
temperature of this blackbody is between 1135 and 1160 K with a
nominal best fit around 1150 K ($\chi^2_{\nu} \approx
0.3$). Fig.~\ref{fig:subfit} shows that this single temperature
blackbody does indeed convincingly reproduce the entire IR-submm
continuum. \\
\begin{figure}
\resizebox{\hsize}{!}{\includegraphics{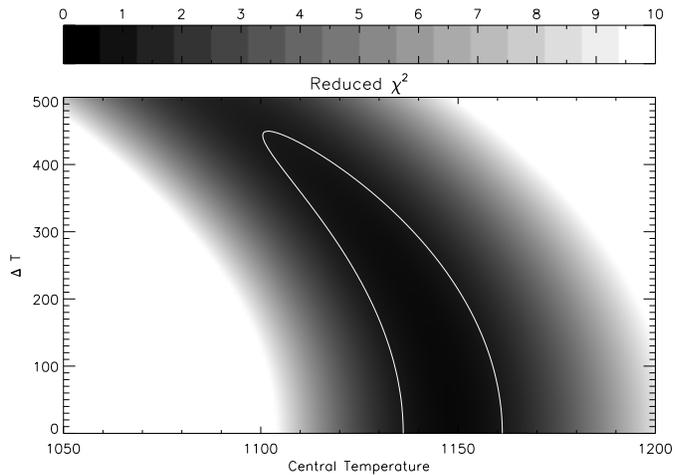}}
\caption{\label{chi2_contours}$\chi^2_{\nu}$ contours for a sum of
  equally weighted blackbody functions. The horizontal axis is the
central temperature in this sum, the vertical axis is the temperature
range around this central temperature; the $\Delta T$ = 0 case are
single temperature blackbodies. The white contour is the $\chi^2_{\nu}
= 1$ level, indicating that such a sum of blackbodies is
indistinguishable from a single temperature blackbody even if the
temperature range is as broad as 400 K. } 
\end{figure}
If the IR emission is really produced by dust grains, it is unlikely
that all the dust is really emitting at a single temperature. Rather,
a distribution of dust centered at a central temperature T$_{c}$
with a small temperature range $\Delta T$ around it would likely
produce a continuum that mimics a single temperature blackbody. 
In order to get a feeling for the allowable
range of temperatures, we therefore performed a second set of $\chi^2$
minimizations where we fitted the observations with a sum of equally
weighted blackbodies with temperatures between $\Tcenter - \frac{\Delta
  T}{2}$ and $\Tcenter + \frac{\Delta T}{2}$. The $\chi^{2}_{\nu}$
contour map is shown in Fig.~\ref{chi2_contours}. As can be seen from
this Figure, the allowable temperature range for such a sum of
blackbodies is 
\begin{equation}
\label{eq:Tmin}
T_{\rm min} \geq 880 K 
\end{equation}
and 
\begin{equation}
\label{eq:Tmax}
T_{\rm max} \leq 1325 K
\end{equation}
corresponding to a temperature range of about 450 K, centered around
1100 K.  Dust grains with a grey opacity and with a radial power-law
number density distribution $N(R) \propto R^{-\alpha}$ would imply that
we have to give weights to the blackbody functions at different
temperatures, with $W(T) \propto T^{2\alpha}$. We also made a $\chi^2$
fit for a $\alpha$=2 and find temperature limits of 730 and 1285\,K.

\subsection{Geometrical constraints}
\label{sec:geom-constr}
The main goal of this paper is to investigate which kind of dust
distribution could reproduce the blackbody shape and the luminosity of
the IR excess. 
\citet{rens:HR4049-dust} actually modeled the IR excess with a
spherical dust shell composed of small dust grains.
This model could reproduce the IR excess down to 100
\micron, showing that a fine-tuned distribution of small dust grains
can result into a blackbody like SED.  However, in order to achieve
the blackbody slope at long wavelengths, the dust distribution must be
very extended since the contribution of low-temperature grains is
needed.  In the model by \cite{rens:HR4049-dust} the distribution
extended out to 2500R$_{\star}$ (corresponding to 500AU) in order to
account for the 60\micron flux.  Since there is now evidence that the
spectrum continues as a blackbody down to at least 850\micron, the
required distribution would have to extend to temperatures as low 3K,
while the interstellar radiation field alone already heats dust to
considerably higher temperatures. We will therefore not investigate
such a fine-tuned distribution of small grains in a spherical shell
any further.  \\
There are two remaining possibilities to explain the
blackbody shape of the IR-excess. The first possibility would be that
the individual grains that are responsible for the continuum emission
are radiating like a blackbody themselves. In that case, the grains
must be large compared to all wavelengths present in the SED, 
and the temperature distribution relatively narrow. 
The second possibility is that the individual grains are not blackbody
emitters; in that case we propose that the blackbody radiation is
produced by an optically thick disk surface which radiates only in a
narrow temperature range. In the remainder of the paper, we will
investigate these two possibilities and their consequences. \\
Assuming that the IR emission is produced by a disk with height H above
the midplane (total thickness of the disk is 2H), the dimensions can
be derived from the constraint that one third of the stellar
luminosity should be absorbed by the disk.  If the disk is infinitely
optically thick to the stellar radiation (like a solid surface) and
located at a distance $R$ from the star, then the disk surface is $2
\pi R\cdot 2H$; this surface must then cover a solid angle of
$\frac{1}{3} \times 4\pi R^2$, so we find for the height of the disk
\begin{equation}
\label{eq:11}
\frac{H}{R} = \frac{1}{3}
\end{equation}
This is actually a lower limit for the height.  If the disk is
not optically thick but absorbs on average only a fraction
$1-e^{{-\tau}}$ of the stellar radiation, the height would have
to be larger by a factor $1/(1-e^{{-\tau}})$. 
\begin{equation}
\label{eq:10}
\frac{H}{R} = \frac{1}{3} \frac{1}{1-e^{{-\tau}}}
\end{equation}
As the only assumption that went into Eqs.~(\ref{eq:11}) and
(\ref{eq:10}) is that the dust is in a disk, the limit on the height
$H$ is a major constraint on any disk model.

\section{Narrow temperature distribution of blackbody grains}
\label{sec:optically_thin}
We will now investigate the possibility that the blackbody shape of
the continuum is due to a collection of ``large'' grains that each
radiate like a blackbody.  ``Large'' grains in this case means that
the size of these grains must be larger than all wavelengths present
in the SED; therefore the grains must be at least mm-sized.

\subsection{Required disk structure}

Using the equation of thermal equilibrium
of dust grains, we can translate the limits imposed on the
temperature from Eqs.~(\ref{eq:Tmin}) and (\ref{eq:Tmax}) into limits
on the distance of the dust grains from the central star :
\begin{equation}\label{eq:tdust-from-star}
T_d=\left(\frac{1}{2}\frac{R_{*}}{R}\right)^{1/2}T_{*}
\comma
\end{equation}
Using a stellar effective temperature $T_{\star}$ of 7500 K and
an effective radius $R_{\star}$ of 47 \Rsun
\citep{bakker:hr4049-variability}, we find that the dust grains must
be between 3.5 and 8 AU from the star. \\
There are a few caveats on these distance limits. Since these numbers
have been calculated assuming that each dust grains receives the
unattenuated radiation from the star, the outer radius is an upper
limit. If we take into account that a fraction of the stellar
radiation gets absorbed, it can be shown that the outer distance
decreases for small optical depth with a factor of $e^{-\tau/2}$;
assuming an optical depth of $\tau=0.5$, the outer limit would
therefore go down to 6.2 AU.  According to Eq.~(\ref{eq:10}) the same
optical depth would require the disk height to reach $2H=1.7R$.  In
this way, the radial extent of the disk can be traded against its
height.  For very low optical depths, the disk must be extremely high.
For modest optical depths, the disk height decreases towards the limit
given in Eq.~(\ref{eq:11}),
but the radial extent must be smaller. \\
We can also estimate the disk mass.  The total absorbing surface is
$A_{\rm abs} = 2\pi R\cdot 2H$ which must be equal to the cross
section of N grains: $2\pi R\cdot 2H = N\pi a^2$.  The total dust mass
is then
\begin{equation}
\label{eq:13}
M_{\rm dust} = 2.85 \cdot\ten{-6}\Msun \frac{R}{4\AU} \frac{H}{1.4\AU}
\frac{\rho_{\rm dust}}{2.7\mathrm{gcm}^{-3}} \frac{a}{\mathrm{1mm}}
\end{equation}
%
%
%
It is clear that the major difficulty of the model with large grains
in a narrow temperature range is how to realize the very particular
arrangement needed; 
\begin{enumerate}
\item  The disk needs to have a fine-tuned radial
optical depth between one and two over all scale heights.  Optical
depths less than one will fail to absorb enough stellar radiation.
Optical depths larger than two will produce a temperature range which
is too wide.
\item\label{item:6}  A disk is required which is more extended in the
vertical then in the radial direction
\end{enumerate}
The scale height of such a disk could be provided either by a gas-rich
disk with a hydrostatic distribution of gas in which the grains are
suspended, or dynamically in a gas poor disk.  We will discuss both
scenarios.
\subsection{Dust grains in a gas-rich disk}

If the disk were gas-rich, the same hydrostatic pressure which is
required to hold up the vertical height would also act in the radial
direction, attempting to spread the disk in this direction.  Since in
a Keplerian disk the radial component of the gravitational force is
balanced by the centrifugal force, stability of the disk would also
require significant sub-Keplerian rotation in the outer parts of the
disk in order to balance the large gas pressure gradient in the disk.
The resulting shear motion leads to transport of angular momentum
which would again spread out the disk in radial direction.
Another concern is grain settling.  The settling time for grains is a
gas-rich disk is given by eq.~(\ref{eq:5}).  If we scale the dust mass
of \cdrev{$2.85 \cdot\ten{-6}\Msun$} with a nominal gas-to-dust ratio of 100,
we can estimate a gas mass in the disk of \cdrev{$2.85 \cdot\ten{-4}\Msun$}.
Using 3.5\,AU and 6.2\,AU for the radial extent of the disk, we find
an average surface density of 30\,g/cm$^2$.  At 4\,AU, this
corresponds to a settling time of 150 yr for 1\,mm dust grains.  Grain
settling would half the disk height in this time scale.  This estimate
is independent of the grain size since the settling time depends on
the ratio of surface density over grain radius which is constant for
the above estimate.  A large height could be kept for a longer time
only if the gas-to-dust ratio in the disk would be much greater than
100.

It is therefore highly questionable if a cloud of large dust grains
supported by a hydrostatic gas-rich disk could reach a stable
configuration in accordance with the observational constraints.

\subsection{Dust grains in a gas-poor disk}
\label{sec:gas-poor}

The other possibility is that the disk is gas-poor.  It
would then be a structure resembling the asteroid belt or the Kuiper
Belt in the solar system, or the debris disks found around
main-sequence stars \citep[e.g.][]{backman-review,jura-zeta-lep}.  In
such disks dust grains are produced by collisions between larger
bodies.  Since grains smaller than about 1mm would be blown away by
radiation pressure, this would also lead to a natural explanation for
the absence of small grains indicated by the blackbody spectrum.  In a
gas-poor disk, the dust grains move on Keplerian orbits around the
central star. The vertical structure of such a disk is then given by
the distribution of orbital inclinations.  One could imagine that the
radial extent of such a disk is set by a sheparding planet, much like
\cdrev{it} is in Saturns rings.

However, it is easy to show that such a disk cannot keep the required
large inclination distribution.  As we have discussed above, the disk
must have an optical depth of order one in the radial direction.
Since the disk is higher than its radial extent, this immediately
implies that the vertical optical depth will be of order one or even
larger.  Therefore, a dust grain moving on an inclined orbit will
experience at least two collisions during each orbit.  This is simply
due to the fact that both the collision cross section and the
absorption cross section are set by the geometrical cross section of
the grains.

The collisions between grains will either destroy the grains if
the velocity is larger than a few km/s \citep{tielens-destruct}, or
dissipate energy.  The energy dissipation leads to a rapid flattening
of the disk.  \citet{jeffreys-saturn} showed for the case of Saturns
rings that such a disk quickly collapses to a mono layer thickness.  A
more detailed study including the effects of gravitational stirring of
bodies embedded in the disk \citep{cuzzi-saturn} showed that the
thickness is given by a few times the diameter of the largest bodies
in the disk.  To counteract the collapse of the disk, massive external
stirring would be needed, which would spread the disk in vertical and
radial direction, in contradiction to the observed narrow range of
temperatures.  

\cdrev{An interesting analogy for this model is the star HD\,98800 which
  was described by \citet{1993ApJ...406L..25Z}.  This star is a binary
  system composed of two pre-main-sequence K dwarfs.  The dust disk,
  located around component B, absorbs about 20\% of the radiation of
  this component.  It is easy to see that the large scale height
  required cannot be due to hydrostatic equilibrium (see section
  \ref{sec:vert-scale-height}) since the stars are not luminous enough.
  Therefore the most likely explanation for this system is that it is
  indeed a debris system.  However, the same arguments limiting the
  lifetime of such a system apply \citep{1993ApJ...406L..25Z}.
  Possibly, the disk in this system has been produced by a recent
  destructive event, or is maintained by the interactions of the
  binary system.}

\section{An optically thick disk: The wall model}
\label{sec:optically_thick}

The second explanation for the blackbody spectrum of HR4049 is that
the emission comes from an optically thick disk surface.

The scenario which we have in mind is a highly optically thick
cylindrical (see Fig.~\ref{fig:sketch}) circumbinary dust disk which
is illuminated from inside by the central star. The high optical depth
prevents radiation to penetrate deeply into the disk which is
therefore cold and invisible against the bright inner rim. What we
observe as the blackbody radiation is the part of the inner disk rim
which is directly visible to the observer.

The disk will have a very hot inner rim, which will be at a single
temperature if the rim is concave (i.e. at constant distance from the
star).  A vertical or more convex inner rim will produce a spectrum
that is a bit more multi-temperature, but this deviation may still lie
within the temperature constraints derived in
Sec.~\ref{sec:temperatures}. 

The optical depth of this disk must be high, at all wavelengths
considered. In fact, the optical depth must be so high that the colder
dust which is not directly illuminated by the star does not contribute
to the emergent spectrum.  We will quantify this below.

In the optically thick model, the constraints on the particle size and
on the radial extent which we derived in
Sect.~\ref{sec:optically_thin} can be relaxed.  The individual
grains do not have to be large, and there is no limit on the radial
extent.




\subsection{Geometrical constraints revisited}
\label{sec:geom-constr-revis}

In Sect.~\ref{sec:geom-constr} we have shown that the height of an
optically thick disk  must be $H=\frac{1}{3}R$ in order to absorb
sufficient radiation, which was derived from the ratio of IR and
stellar luminosities.

\begin{figure*}
\includegraphics[width=\textwidth]{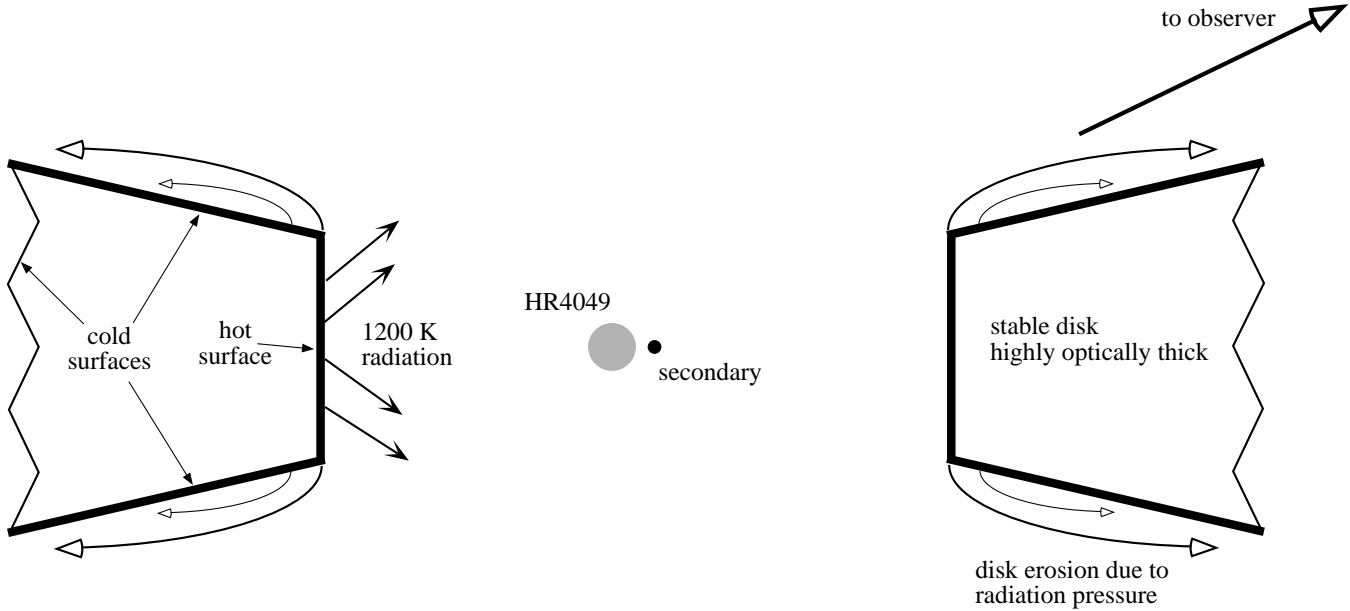}
\caption{Geometry sketch of the wall model for HR4049.  The double
  star is at the center of the fat disk.  The inner surface of the disk
  is heated by the star and produces the blackbody radiation.  The
  material behind the inner surface is shielded and cold, including the top,
  bottom and rear surfaces. The height of the disk is set by
  strong radiation forces which blow away the upper parts of the
  disk.  The direction to the observer is such that we look through
  the edge of the disk.}
\label{fig:sketch}
\end{figure*}

We envision the emitting surface to be optically thick and
radiating only into a half-space (see Fig.~\ref{fig:sketch}).  The
flux seen by an observer will depend upon the viewing angle under
which the system is seen.  From a nearly pole-on direction, the
projected surface would be small and the IR radiation received low.
From a nearly edge-on direction, the projected surface is large and
the observed flux high. The observed IR luminosity therefore roughly
scales with $\sin i$. Since the $H/R$ ratio depends on the emitted
IR luminosity rather than the observed luminosity, there is an
uncertainty in the real value of $H/R$, depending on the inclination
at which we observe the system.   
%
%
Because of the significant variations in circumstellar extinction
during the orbital motion of HR4049, a larger (near edge-on)
inclination seems more likely, meaning that the real $H/R$ could be
somewhat smaller than $\frac{1}{3}$. \\ 
From the temperature of the dust one can easily compute the distance
of the inner rim to the central star.  This distance is a factor of
two larger than the distance of black body grains in an optically thin
environment because the surface can cool only into a solid angle of
$\pi$ instead of $4\pi$ in the optically thin case.  If this rim is
located at a distance $R$ from the star, then the stellar flux it
receives is: $F_{*}=(R_{*}/R)^2\sigma T_{*}^4$. The flux emitted by
the wall is $F_{\mathrm{wall}}=\sigma T^4$. Therefore the temperature
of the wall, if located at a distance $R$, is
$T=(R_{*}/R)^{1/2}T_{*}$.  For $T=1200\kel$, $T_{*}=7500\kel$ and
$R_{*}=47 R_{\odot}$ one obtains a radius of the inner surface of
$8.5\,\AU$. If one considers the fact that the wall is also
irradiating itself, then one must take into account that the total
irradiation flux of the wall will be about a factor of $1+\Omega/4\pi$
higher than the direct stellar irradiation. By taking
$\Omega/4\pi\simeq 1/3$, as derived above, one finds
$R=\sqrt{4/3}\times 8.5\,\AU =9.9\,\AU$. It should be noted that the
estimate of $1+\Omega/4\pi$ as the enhancement factor for the
irradiation is very rough, but suffices for now.  For a detailed
discussion, see \citet{DDN}.

\subsection{Emission from an irradiated dust wall}

We now turn to the question of whether an irradiated wall will indeed
produce an emission that is sufficiently close to a blackbody curve to
fit the SED of HR4049.  \cdrev{Since 2D radiative transfer calculations
  with full frequency dependence and at optical depths reaching a
  million are very cumbersome, we will tackle this problem in a
  two-step approach.}: First we will use a 1-D radiative transfer
calculation in order to see if the inner surface of the wall does
indeed produce blackbody-like emission.  Then we will use a 2-D
diffusion calculation in order to see if the emission of the rest of
the disk (surface and outer rim) can be sufficiently suppressed using
high optical depth.

For the 1-D calculation, we consider infinite half-space filled with
dust: for $x<0$ there is vacuum, for $x>0$ there is a constant density
medium of dust grains.  This half-space is illuminated (on the $x<0$
side) with a flux $F_{\nu}^{*}$ which is taken to be the stellar flux
as observed at a radius of $8.5\,\AU$ (we ignore the effects of
self-irradiation mentioned above).
We run the calculation for a mixture of grains of two sizes: small
(0.1\micron) grains, and large grains.  We take the opacity of
\citet{draalee84} for small grains, and a simple grey opacity for our
'large' grains. \cdrev{The grey opacity guarantees that the optical
  depth will be large at all wavelengths, a very important requirement
  of the wall model.  We include the small grains in order to see if
  the superheating of small grains close to the surface of the disk
  will cause significant deviations from the blackbody shape.  The
  requirement of large optical depth at all wavelengths could also be
  met with non-grey, small-grain opacities.  In this case the limiting
  criterion would be that the optical depth at 850 \micron must be
  large.}  The transfer is performed using a frequency-dependent
discrete-ordinate method (i.e.~ray-by-ray transfer at each frequency
bin and for each angle). The transfer equation is:
\begin{equation}
\mu\frac{d I_{\mu,\nu}}{dx} = \sum_i \rho_i\kappa^i_\nu [ B_\nu(T_i)
- I_{\mu,\nu} ]
\end{equation}
where $\kappa^1_\nu$ and $\kappa^2_\nu$ are the opacities of the small
grains and the large grains respectively, and idem for the densities $\rho_1$
and $\rho_2$ and the temperatures $T_1$ and $T_2$. We ignore scattering for
simplicity. The symbol $\mu$ is the cosine of the angle of the radiation. At
every point $x$ in the slab, the temperature of the dust grains is
determined by the radiative equilibrium equation:
\begin{equation}
\int_0^\infty \kappa_\nu B_\nu(T_i) d\nu 
= \int_0^\infty \kappa_\nu J_\nu d\nu 
\end{equation}
where the mean intensity $J_\nu$ is given by
\begin{equation}
J_\nu = \frac{1}{2}\int_{-1}^{1} I_{\mu,\nu} d\mu
\end{equation}

\begin{figure}
\includegraphics[width=8.5cm]{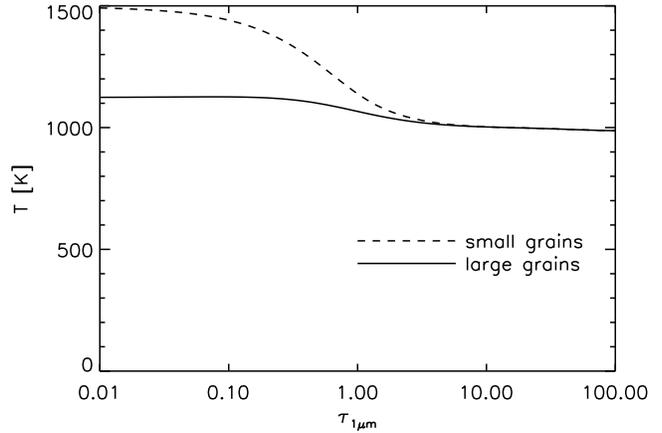}
\caption{\label{fig:slab-temp} Temperature structure of the 1D calculation.}
\end{figure}
\begin{figure}
\includegraphics[width=8.5cm]{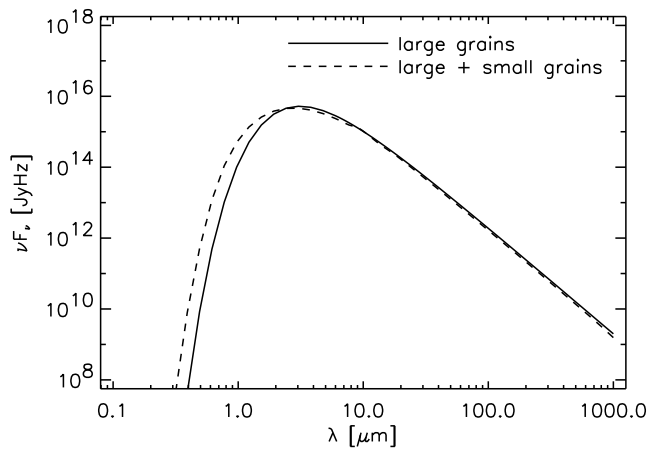}
\caption{\label{fig:slab-sed} The SED produced by the 1D calculation.
  The solid curve is indistinguishable from a single temperature
  blackbody.}
\end{figure}

Using Lambda Iteration with Ng acceleration we iterate this set of equations
to convergence and find the solutions shown in Fig.~\ref{fig:slab-temp}. 

One sees that for the large grains, the temperature remains almost
constant as a function of optical depth.  The small grains superheat
on the disk surface.  At higher optical depth, the two temperatures
become the same because at large optical depth both grain species
become radiatively coupled.  By performing direct ray-tracing through
this model atmosphere, we can obtain the predicted SEDs, which are
shown in Fig.~\ref{fig:slab-sed}.  The emerging spectra
are close to pure blackbody curves, in fact the fat curve is
indistinguishable from a blackbody. The superheated small grains
produce a slight broadening of the SED towards short wavelength.  We
conclude that an irradiated high optical depth surface indeed radiates
like a blackbody and could in principle be used to explain the SED of
HR4049.

\subsection{Diffusion though the disk}
\label{sec:diffusion}
In order to address the issue of diffusion through the disk towards the
surface and towards the outer edge, we need to perform a 2-D radiative
transfer calculation. Unfortunately, realistic radiative transfer
calculations in 2-D are rather cumbersome and complex, in particular at the
very high optical depths needed for our disk model. Under these
circumstances, a reasonably reliable approximation of the 2-D transfer
problem can be made using the equations of radiative diffusion.

Since we are interested in the qualitative behavior of the radiative
diffusion through the disk, we keep the disk geometry highly simplified. We
assume a constant density torus limited in radius by $9\AU<R<18\AU$ and in
$\Theta$ (the angle from the pole) between $\pi/2-1/3<\Theta<\pi/2+1/3$ -
consistent with the $H/R=1/3$ requirement. We compute first how the direct
stellar radiation penetrates the inner few mean free paths of the inner
edge:
\begin{equation}
F^{*}_\nu(R,\Theta)=\frac{L_\nu^{*}}{4\pi R^2}\;
\exp\left\{-\int_{\rin}^{R}\rho\kappa_\nu dR\right\}
\end{equation}
We now assume that the absorbed radiation is re-emitted as IR radiation.
This diffuse IR radiation is treated as a separate radiation field. We do
this in a frequency-averaged way, and write down the diffusion equation for
the frequency-averaged mean intensity $J(R,\Theta)$:
\begin{equation}\label{eq-diff-twodee-eq}
\begin{split}
\frac{1}{R^2}\frac{\partial}{\partial R}
\left(\frac{R^2}{3\rho\kappa}\frac{\partial J}{\partial R}\right) &+
\frac{1}{R\sin\Theta}\frac{\partial}{\partial\Theta}
\left(\frac{\sin\Theta}{3 R \rho \kappa}\frac{\partial J}{\partial
\Theta}\right) = - S
\end{split}
\end{equation}
where $\kappa$ is the Rosseland mean opacity. This equation describes the
radiative diffusion in polar coordinates, derived from the first two moment
equations with Eddington closure. We also implicitly assume (for simplicity)
that the mean opacity does not change very much from location to location,
since otherwise additional terms have to be included in the diffusion 
equation.

The source term $S$ is given by:
\begin{equation}
S=\frac{1}{4\pi}\int_0^\infty \rho \kappa_\nu F^{*}_\nu d\nu
\end{equation}
This source term describes the transition of energy from primary stellar
radiation into reprocessed diffuse infrared radiation. Since all primary
radiation is absorbed in the innermost few mean free pathlengths, this
source term will be only non-negligible close to $R=\rin$.  So for larger
radii the above diffusion equation (\ref{eq-diff-twodee-eq}) effectively
states that the divergence of the flux should be zero, i.e.~that energy is
conserved. The boundary conditions at all edges are taken to be those
described in \citet{rybickilightman:1979}.

The 2-D diffusion equation can now be solved by writing it into a matrix
equation, and solving this using standard techniques such as the biconjugate
gradient method (e.g.~Numerical Recipes \citeyear{numrecip:1992}). Once we
find $J(R,\Theta)$, we can compute the dust temperatures from:
\begin{equation}
\frac{\sigma}{\pi} T(R,\Theta)^4 = J(R,\Theta) + \frac{S}{\rho\kappa_P(T)}
\end{equation}
where $\kappa_P(T)$ is the Planck mean opacity at the dust temperature. 

\begin{figure}
\includegraphics[width=8.5cm]{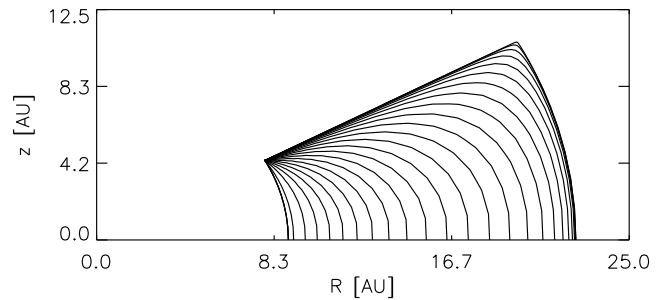}
\caption{\label{fig:diffusetemp} Temperature contours of the diffusive
  disk.  The temperature at the inner boundary is 1200K, further
  contours are in steps of 50K.}
\end{figure}

\begin{figure}
\includegraphics[width=9.0cm]{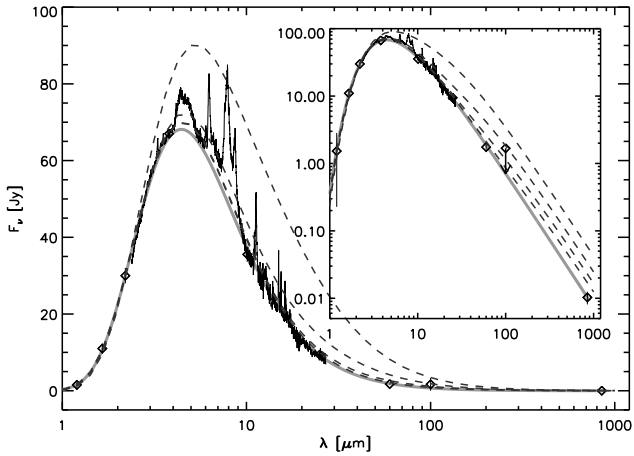}
\caption{\label{fig:diffusefit} Fits to the observed SED with
  diffusion calculations for tori with different optical depth.  The
  solid grey line shows the 1150K blackbody fit.  The dashed lines
  are (from top to bottom) for optical depth 10$^2$, 10$^3$, 10$^4$,
  and 10$^5$.}
\end{figure}

We run several models, with an equatorial Rosseland optical depth between
$\tau=100$ and $\tau=10^5$ and one for $\tau=10^6$.  The resulting
temperature profile for the $\tau=10^4$ case is shown in
Fig.~\ref{fig:diffusetemp}.  In Fig.~\ref{fig:diffusefit} we show the SEDs
of these tori, as seen from an inclination of 45 degrees. We used a 2-D
ray-tracing utility to produce these SEDs from the temperature distribution
resulting from the 2-D diffusion equations.

One sees that even for very high optical depth the spectrum will
exceed a pure blackbody curve at long wavelengths.  The $\chi_{\nu}^{2}$ for
the $\tau=10^5$ fit is 2.85.  For high optical depths only a very tiny
amount of radiation is leaking through the disk. But since the
Rayleigh-Jeans slope of a blackbody curve does not contain much
energy, even this tiny amount of leakage shows up at long wavelengths.
These results are in agreement with the simple estimate that the flux
at the surface and the outer edge of the torus is proportional to
$1/\tau$, where $\tau$ is the optical depth towards the inner edge. If
we assume that the flux is proportional to $T^4$, then one can
estimate that the temperature of the outer- and upper/lower surfaces
of the torus is proportional to $T_{\mathrm{surf}}\propto
\tau^{-1/4}$. This means that in order to suppress the temperature of
these back-surfaces of the disk by a factor of $0.1$ one must have an
optical depth of $\tau\simeq 10^4$.  A pure blackbody emission from
this torus model therefore requires very high optical depth.

\section{Disk properties}
\label{sec:disk-properties}

The disk we are proposing to be the source of the 1150\,K blackbody
radiation in HR4049 certainly is unusual, because of the high scale
height (covering fraction) at the inner boundary and the high optical
depth. In this Section, we will derive properties of the disk including
mass estimates and stability considerations in order judge how
realistic such a model can be. We will therefore investigate the
following questions. 

\begin{enumerate}
\item Can we understand the large disk height ($H/R=1/3$)?
\item\label{item:4} Can we place limits on the grain size?
\item\label{item:1} What disk mass is required to make the disk
  sufficiently optically thick?
\item\label{item:2} How long will it take for the dust grains to
  settle towards the midplane?
\item\label{item:5} Will the disk remain radially mixed, or will the
  radiation pressure of the star lead to a grain-gas separation?
\item\label{item:3} Will radiation pressure of the star be sufficient
  to blow away (parts of) the disk?
\end{enumerate}

As we already discussed in Sect.~\ref{sec:gas-poor}, a gas-poor disk
cannot keep the required scale height because of frequent collisions.
This argument is even much stronger for an optically thick disk. We
will therefore assume the disk is gas-rich and in hydrostatic
equilibrium, to which we can apply the standard theory of hydrostatic
disks.  We will adopt a cylindrical axisymmetric coordinate system
with radial coordinate $r$ and vertical coordinate $z$. We will often
refer to scale heights in the following sections; we remind the reader
that these scale heights correspond to the vertical coordinate
starting from the equatorial plane.  For the following estimates we
will for simplicity assume that the disk has a constant surface
density $\Sigma_0$ and a constant scale height $\hnull$ \cdrev{(i.e. a
  non-flaring disk, see below)} in a distance
range $\Delta r=r_1-r_0$ from the star.
We will further assume that inside the
disk the dust-to-gas ratio is $f_{\dust}=0.01$ and that the vertical
distribution of the gas is given by hydrostatic equilibrium in an
isothermal disk.  We assume the dust particles all to have the same
size $a$ and a specific density $\rho_{\dust}$.

Armed with these properties, we may calculate the total disk mass
\begin{equation}
\label{eq:1}
M_{\disk} = 2\pi (r_1^2-r_0^2) \Sigma_0 \fullstop
\end{equation}

When evaluating equations numerically in the following, we will always
use a standard model of the disk which contains 0.3M$_{\odot}$, with
$r_0=10$\,AU and $r_1=20$\,AU.  This corresponds to a surface density
of 1414\,g\,cm$^{-3}$.  \cdrev{The disk mass of 0.3M$_{\odot}$ is very
  high and should be considered an upper limit for the possible disk
  mass since significantly higher masses will lead to gravitational
  instabilities if $M_{\rm disk}>\frac{H}{R}M_{\star}$ \citep{1981ARA&A..19..137P}.  We use this high mass in order to
  maximize the stability of the disk against grain settling or
  radiation pressure.}

\subsection{Vertical scale height from \cdrev{hydrostatic equilibrium}}
\label{sec:vert-scale-height}

\cdrev{We assume the disk to be gas-rich.  Its height can therefore be
  calculated from the assumption of hydrostatic equilibrium}.
The disk pressure scale height $\hp$ is given by the square root of
the ratio of the sound speed and the local Kepler frequency and can be
written as \citep[e.g.][]{CG97}

\begin{equation}\label{eq:8}
\frac{\hp}{r} = \chi \sqrt{\frac{k T r}{G (M_1+M_2)\mu m_p}} 
\end{equation}
where $M_1$ and $M_2$ are the masses of the two stars in the HR4049
system, $\mu$ is the mean molecular weight and $m_p$ is the proton mass.
It should be noticed that this formula holds for an annulus of a thin disk, 
and may be inaccurate for the inner edge of a thick disk.  
For $r=8.6\AU$, $M_1\simeq 0.56\;M_{\odot}$ and $M_2\simeq
 M_1$, and $T=1150$\,K this amounts to
\begin{equation}
\label{eq:pressure-scaleheight}
\frac{\hp}{r} = 0.19
\end{equation}
If we compare this computed scale height with the observed required
scale height (see Sect.~\ref{sec:geom-constr}), which was $\hs/r=1/3$,
and if we take into account that the surface height (the height at
which $\tau=1$) is usually a few times the pressure scale height we
arrive at
\begin{equation}
\frac{\hs}{r} = 0.19 \frac{\hs}{\hp}
\end{equation}
where $\hs/\hp$ is roughly in between $1$ and $4$ for a Gaussian 
vertical density distribution \citep{CG97}

Therefore, a hydrostatic disk can reach the required height
without any difficulty.  In fact, a mechanism to truncate the disk
somewhere between one and two pressure scale heights seems to be
required in order to match the observational constrains.

\cdrev{In the passive disks around young stars, the surface
  irradiation of thin circumstellar disks leads to an increase of the
  ratio $\hp/r$ with distance from the star.  Such disks are called
  \emph{flaring disks}, and have been successfully used to explain the
  flat SEDs of young stars with disks
  \citep{1987ApJ...323..714K,CG97}.  Could the disk around HR4049 also
  be flaring?  The blackbody shape of the SED is clear evidence that
  this is not the case, because flaring disks generally have very
  broad, multicomponent SEDs.  The physical criterion for flaring to
  occure is that the disk midplane temperature must decrease more
  slowly than $r^{-1}$, so that $\hp/r$ increases with distance (see
  eq.~(\ref{eq:8})).  In the wall model, the disk temperature is not
  determined by the radiation reaching the disk surface.  Instead,
  radiation is only impacting on the inner rim and transported
  outwards through the disk by diffusion.  The temperature will
  therefore decrease quickly with distance from the star, and $\hp/r$
  will decrease, preventing any flaring.  The shape of such a diffuse
  disk inner rim has recently been discussed in detail in connection
  with the near IR emission of Herbig AeBe stars \citep{DDN}.}


\subsection{Radial optical depth}
\label{sec:radial-tau}
\begin{figure}
\includegraphics[width=8.5cm]{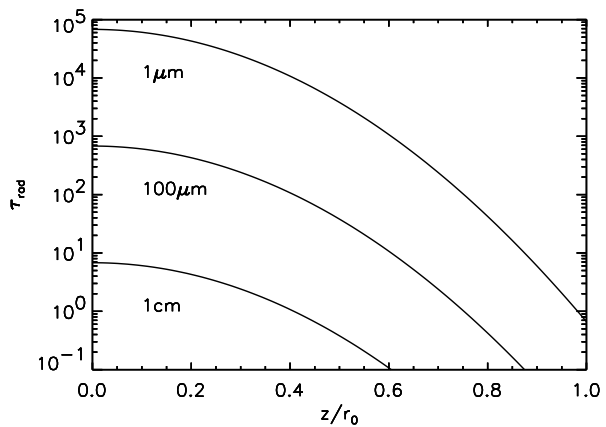}
\caption{\label{fig:taurad} The optical depth for short wavelength
  radiation along radial rays through the torus as a function of
  height.  The different curves are for different grains sizes with
  fixed total mass of 2M$_{\odot}$.}
\end{figure}

The density structure $\rho(z)$ in the disk assuming hydrostatic
equilibrium is given by \citep{CG97}
\begin{equation}
\label{eq:2}
\rho(z) = \frac{\Sigma_0}{\sqrt{2\pi}\hnull} \exp \left\{ -\frac{z^2}{2\hnull^2}
\right\}
\fullstop
\end{equation}
If dust and gas are perfectly mixed, the number density of dust grains
is given by
\begin{equation}
\label{eq:3}
n_{\dust}(z) = \frac{f_{\dust}\rho(z)}{m_{\dust}}
= \frac{3 f_{\dust}\rho(z)}{4\pi a^3 \rho_{\dust}}
\end{equation}
where $f_d$ is the dust-to-gas ratio and $m_{\dust}$ is the mass of a
single dust grain.  The dust opacity is then given by
\begin{equation}
\label{eq:4}
c_{\dust}(z) = n_d Q(a,\lambda) \pi a^2 = \frac{3 f_{\dust}\rho(z)Q(a,\lambda)}{4 a \rho_{\dust}}
\end{equation}
where $Q(a,\lambda)$ is the absorption efficiency factor.
To calculate the radial optical depth we use the assumption, that the
radial extent of the disk is small compared to the distance of the
inner rim to the star, so that we can consider the ray going through
the disk material on a constant height $z$.  The optical depth on such
a ray is given by
\begin{equation}
\label{eq:6}
\tau(z,\lambda) = c_{\dust} \Delta r 
= \frac{3 f_{\dust}\Delta r \Sigma_0 Q(a,\lambda)}{4 \rho_{\dust} a
  \sqrt{2\pi}\hnull}
\exp \left\{ - \frac{z^2}{2\hnull^2} \right\} 
\fullstop
\end{equation}

In Fig.~\ref{fig:taurad} we show the optical depth for our standard
disk as a function of height $z/r_0$ in the disk.  Since we keep the
total disk mass constant, the optical depth is different for different
grain sizes.  What we have plotted is the optical depth at very short
wavelength, using a geometrical cross section for the grains.  This
number is an estimate for the optical depth at all wavelengths smaller
than $2\pi a$.  For interstellar grain sizes (0.1\micron), very high
optical depths of up to 10$^6$ can be reached, but only at short
wavelengths.  The SED measurement with the longest wavelengths is the
850\micron SCUBA measurement, and the 100\micron grains show a better
estimate for the optical depth at these wavelength.  Here the standard
disk model can reach $\tau=10^3$, and with a disk mass of
$2M_{\odot}$, $\tau=10^4$ might be feasible.  As shown in
Fig.~\ref{fig:diffusefit}, this optical depth gives a reasonable fit
to the observed SED even though the flux at 60\micron and 850\micron
is still slightly over predicted.  We therefore conclude that a massive
disk can indeed reach very high optical depth. But we also find that
the grains should not be much bigger than 100\micron, or the optical
depth will become too low.

\begin{figure}
\includegraphics[width=8.5cm]{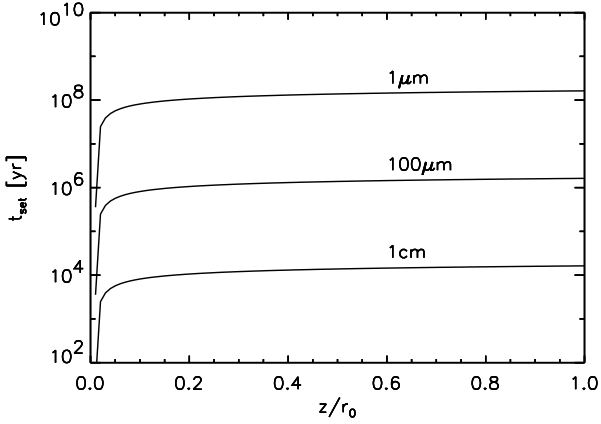}
\caption{\label{fig:settling} Settling times for dust grains of
  different size. }
\end{figure}

\subsection{Grain settling}
\label{sec:settling}

In order to reach the required covering faction, the optical depth of
the disk must be high not only in the midplane, but also at higher z,
up to $z/r_0\approx0.3$.  Since the disk opacity is entirely due to
dust grains, but the disk scale height is entirely due to gas
pressure, gas and dust have to be well mixed.  From protoplanetary
accretion disks it is known that dust grains will eventually start to
settle towards the midplane.
The time for a dust grain to settle from height $z$ to height $z_0$
in a quiescent hydrostatic disk is given by \citet{settling}
\begin{equation}
\label{eq:5}
t_{\settl} = \frac{\pi}{2} \frac{\Sigma_0}{\rho_{\dust}a}
\frac{1}{\Omega_{\kepler}} \ln \frac{z}{z_0}
\end{equation}
where $\Omega_{\kepler}=\sqrt{G M/r^3}$ is the Kepler frequency in the
disk.  

Figure~\ref{fig:settling} shows the settling times for different grain
sizes in the standard disk.  The timescale varies from 10$^8$ years
for 1\micron grains to 10$^4$ years for 1\,cm grains.  Therefore, at least
for grains smaller than about 100\micron, grain settling does not provide
a serious constrain to the disk model.  However, for grains larger
than a few cm, the lifetime of the disk becomes so short that we would
be lucky to see HR4049 in this stage at all.

\begin{figure}
\includegraphics[width=8.5cm]{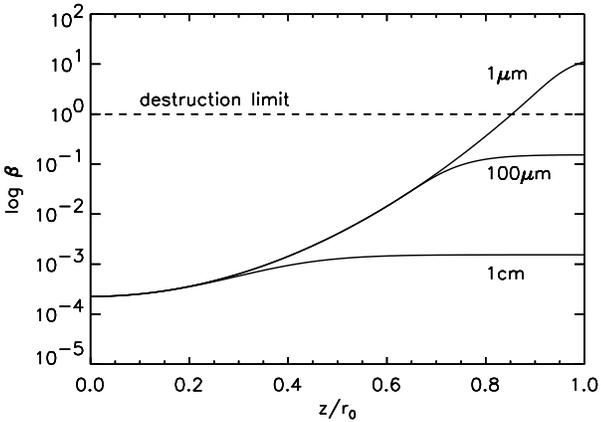}
\caption{\label{fig:blowout} Ratio of total radiation pressure on a
  radial column of material over the gravitational attraction from the
  binary.  Different curves are for different grain sizes with a fixed
  total mass of 0.3M$_{\odot}$.  The dashed line marks the blowout limit.}
\end{figure}

\subsection{Radiation pressure and disk stability}

To estimate at what height the radiation pressure will actually be
large enough to blow away the entire disk (gas and dust), we make the
following calculation.  We consider a radial ``beam'' of material in
the disk.  The beam is all the material outside a surface element
$\Delta S$ in the disk wall.  The radiation pressure available for
that beam of material is given by the momentum flux of all the
radiation entering $\Delta S$ and being absorbed in the beam, thus
$L_{\star}\Delta S(1-e^{-\tau})/4\pi r_0^2 c$. In order to blow away a part
of the disk, the radiation pressure must be strong enough to lift the
entire mass outside of the surface element of the rim.  This mass is
given by $\rho(z) \Delta r \Delta S$.  The ratio of radiative and
gravitational force is then
\begin{equation}
\label{eq:14}
\beta = \frac{L_{\star} (1-e^{-\tau})}{4\pi G M  c \rho(z)\Delta r }
\comma
\end{equation}
and $\beta=1$ marks the limit where the radiation force can remove the
entire beam of material.
Inserting the expression for the density Eq.~(\ref{eq:2}) and solving
for $z$ under the assumption $\tau\gg1$, we find that the critical
height $z_{\rm crit}$ where the disk is
blown away by the stellar radiation  is given by
\begin{equation}
\label{eq:9}
z_{\rm crit}^2 =-2\hnull^2 \ln \left[ \frac{L_{\star}}{4\pi c}\frac{1}{GM}
\frac{\sqrt{2\pi} \hnull}{\Sigma_0 \Delta r} \right]
\end{equation}

Fig.~\ref{fig:blowout} shows the value of $\beta$ for the standard
disk and 3 different grain sizes as a function of height in the
disk. 
The positions where the curves cross the line $\beta=1$ is the
location of $z_{\rm crit}$.  We can see that for grains larger than
$\approx10\micron$, the radiation pressure is never large enough to remove
the disk (gas+dust).  For low height this is due to the large mass in
a radial beam of material.  For large height, the disk becomes
optically thin, and an insufficient fraction of the radiation is
absorbed in the disk.  Only if the dust in the disk consists mainly of
grains smaller than 10\micron, the parts of the disk with $z/r_0>0.8$ can
be completely removed by radiation pressure.

\begin{figure}
\includegraphics[width=8.5cm]{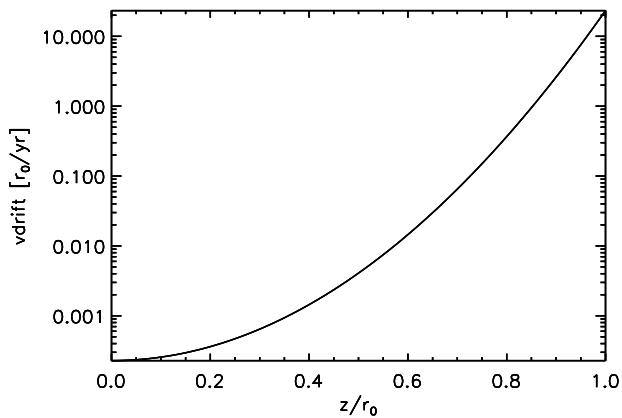}
\caption{\label{fig:vdrift} Equilibrium drift velocity of grains
  assuming direct exposure to unattenuated stellar light.  Note that the
  drift velocity is given in units $r_0$ per year.}
\end{figure}

\subsection{Grain drift}
\label{sec:grain-drift}
Removing a part of the disk is not the only effect radiation pressure
on dust grains can have.  Since the radiation pressure only acts on
the dust grains, the grains can be pushed outwards through the gaseous
disk.  In parts of the disk which are optically thick, all but the
innermost grains are shielded from direct stellar radiation.  The
internal radiation field in the disk is almost perfectly isotropic,
and the radiation force and corresponding drift motion will be small.
However, in the optically thin parts of the disk and also at the inner
rim, grains do receive the full stellar radiation.  In these
regions,  the velocity at which the
grains move can be calculated by assuming an equilibrium between
the radiation pressure on a dust grain and the drag force due to the
motion of the grain.  In the case of subsonic drift, the equilibrium
velocity is given by \citep[e.g.][]{sizedist}
\begin{equation}
\label{eq:7}
v_{\drift} = \frac{L}{4\pi r^2 c} \frac{Q}{\Omega_{\kepler}}
\frac{c_s}{v_{\rm th}} \frac{\sqrt{2\pi}}{\Sigma_0} 
\exp \left\{ \frac{z^2}{2H^2} \right\}   
\end{equation}
In Fig.~\ref{fig:vdrift} we have plotted the drift velocity as a
function of height in the disk.  In the disk midplane, the drift
velocity is about 10$^{-4}$r$_0$/yr.  Grains can therefore be pushed
into the disk on a timescale of 10$^4$ years only.  This defines 
the tightest constraint on the life time of the disk.  Note,
however, that this limit was calculated under the assumption of a
quiescent disk.  Turbulent motions near the inner disk boundary could
be efficient in keeping the grains close to the inner boundary.  At
larger height in the disk, the drift becomes very efficient. At
$z/r_0=1/3$, grains can be pushed through the disk in less than
10$^3$ years.  This provides an effective truncation mechanism for the
disk height and may explain the difference between the calculated
hydrostatic height of the gas disk and the observed covering fraction.

\section{Discussion}
\label{sec:discussion}

The circumstellar environment of HR 4049 is certainly unusual in many
ways.
There are three major constraints which are given by the observations,
and which must be explained by any model:
\begin{enumerate}
\item\label{item:7} The UV extinction is highly variable, non-grey and
  in phase with the orbital motion.
\item\label{item:8} The IR excess contains 1/3 of the stellar
  luminosity.
\item\label{item:9} The spectral shape of the excess is a perfect
  blackbody from 1\micron to 850\micron.
\end{enumerate}

From (\ref{item:7}) one can immediately conclude that the distribution
of material is not spherical.  Our line-of-sight passes through a
region with a strong gradient in extinction, probably the upper part
of a disk-like structure.  The grains causing the extinction must be
small since large grains would imply grey extinction.  If a disk-like
geometry is causing the reprocessing of stellar light, (2) implies
that the disk must be at least marginally optically thick and that it
must have a large height.  The most difficult part turns out to
be explaining (3).  We have presented two different scenarios on how a
disk may produce a blackbody emission: (i) large grains in a
marginally optically thick disk with very small radial extent and (ii)
grains of unknown size in a disk with very high optical depth.  In
this discussion we will highlight the virtues and the shortcomings of
both models.

\subsection{Large grains versus small grains}

If the blackbody shape of the IR excess is caused by thermal emission
of large grains (see Sect.~\ref{sec:optically_thin}), these grains
can only be distributed in a disk. 
%
This disk must have a
large height and it must be confined to at most 10 AU. As discussed in
Sect.~\ref{sec:gas-poor}, it is unlikely that this disk would be a
gas-poor disk. Moreover, the presence of strong molecular bands in the
ISO-SWS spectrum of HR 4049 (see Paper I) strongly suggests that the
circumstellar environment of HR 4049 is indeed gas-rich. The gas
pressure would then support the disk to keep the large scale
height. The main issue to address is then : {\em where are the
  small grains ?}  Some small grains are still required for explaining
the UV absorption as well as the features due to nano-diamonds and
PAHs in the ISO-SWS spectrum.  But there must be an efficient mechanism
to concentrate the small grains in a small region through which our
line-of-sight passes, while effectively removing them from the regions
where most of the reprocessing of stellar radiation happens. 

One way to explain this would be to assume that the small grains are a
completely separate component which is \emph{not} located in the disk, but
above the disk or in bipolar lobes created by a more recent mass-loss
event. This dichotomy between large and small grains is actually
consistent with the conclusions by \citet{joshi:hr4049-polarization}
who argued that the polarization they observed points to two
populations of dust grains, one being very small grains, the other
being large grains. However,  such a geometry makes it more difficult
to understand the large variations of extinction in phase with orbital
variations.

As the gas which is observed through the molecular bands is typical
for an O-rich environment (see Paper I), it is reasonable to assume
that the dust grains which reside in this gaseous reservoir are also
O-rich. If this is true, the complete lack of silicate features in the
spectrum is another indication for the absence of small grains in this
disk. These O-rich dust grains would then incorporate the refractory
elements lacking in the photosphere of HR 4049. \\
If there are mainly large grains in the disk, the dust mass as
derived from Eq.~(\ref{eq:13}) should be a fair estimate for the total
dust mass. Adopting a standard dust-to-gas ratio then yields a total
disk mass of a few times \ten{-4} \Msun. This is somewhat low for the
total amount of mass typically lost on the AGB. As discussed in this
paper, it seems unlikely that a large fraction of the dust has escaped
the system. Moreover, there is no clear-cut evidence (yet) for any
kind of outflow. The rather low total disk mass would then indicate
that the binary system prevented HR 4049 to have a ``normal''
evolution and leave the AGB prematurely, before the star could
experience the most dramatic episodes of mass loss. \\
Apart from explaining the location of the small grains, the main
problem with the large-grain model appears to be how to support the
required disk height for an extended period of time.  Settling (in a
gas-rich disk) and collisions (in a gas-poor disk) both limit the
lifetime severely.

The wall model is an appealing alternative to explain the blackbody
shape of the IR continuum. In this case, a standard grain size
distribution could be present in the circumbinary disk.  The blackbody
emission would be due to the entire surface instead of the individual
grains.  The visible grains near the surface all do have high
temperatures of about 1200K, and the absence of silicate features
would be due to the absence of grains in the right temperature regime,
and to the small temperature gradient near the surface.  The
variations observed in the UV absorption - due to small grains - can
be tied to the vertical density distribution.  Indeed, as explained in
Sect.~\ref{sec:settling}, the differential settling times for
different grain sizes would cause a layered grain size stratification
in which any large grains would settle towards the midplane, while the
smallest grains would still be present high in the disk, causing the
extinction in a beam passing through this region. If we assume that
the vertical density distribution of the small grains is not too
affected by the settling mechanism, the density distribution of these
small grains would follow the gas density as given in
Eq.~(\ref{eq:2}). As the small change in viewing angle between
photometric minimum and maximum causes a large change in the amount of
absorbed energy, we probably view the system with a rather high
inclination, skimming just past the edge of the
disk. \\

\subsection{The inclination of HR4049}

We can actually get a more quantitative idea about the inclination at
which we view the system by expressing it in terms of the hydrostatic
scale height. From Eq.~(\ref{eq:2}) it follows that
\begin{equation}
\label{eq:height}
\frac{d\ln \rho}{d\ln z} = -\frac{z^2}{\hnull^2}
\end{equation}
To calculate $d\ln z$ and $d\ln \rho$, we consider the following. 
At photometric minimum, the star is at inferior
conjunction (i.e. closest to us, poor observers). The line of sight
then crosses the disk at height $H$, thus probing material with a
density $\rho_{\rm min}$. 
At photometric maximum, the star is further away
from us, and therefore the line of sight crosses the disk at
$H+\Delta H$, this time probing material with a density $\rho_{\rm max}$. 
From simple geometric considerations it then follows that 
\begin{equation}
\label{eq:inclination}
\frac{\Delta H}{H} \approx \frac{D}{R_0}
\end{equation}
with $D$ the distance between the star at photometric minimum and at
photometric maximum and $R_0$ the distance between the center of mass
of the binary system and the disk. At the same time, the densities are
proportional to the absorption optical depth so that 
\begin{equation}
\frac{\rho_{\rm min}}{\rho_{\rm max}} = \frac{\tau_{\rm
    min}}{\tau_{\rm max}}
\end{equation}
and
\begin{equation}
\frac{d\ln \rho}{d\ln z} = \frac{d\rho / \rho}{dz / z} = 
\frac{\rho_{\rm max} - \rho_{\rm min}}{\rho_{\rm min}} \frac{H}{\Delta H} =
\left(\frac{\tau_{\rm max}}{\tau_{\rm min}} - 1\right)\frac{R_0}{D}
\end{equation}
Finally, Eq.~(\ref{eq:height}) becomes
\begin{equation}
\frac{z}{\hnull} = \sqrt{\frac{R_0}{D} \left(1 - \frac{\tau_{\rm
        max}}{\tau_{\rm min}}\right)}
\end{equation}
Using the ``grey'' optical depths as derived from the UV absorption (see
Sect.~\ref{sec:energy}), $R_0$=10\AU (as in our standard model in
Sect.~\ref{sec:disk-properties}) and $D \sim $1.18 \AU
\citep{hans:post-AGB-binaries} we find that $z
\sim 2.8 \hnull$.  With $\hnull$ the pressure scale height
(Eq.~(\ref{eq:pressure-scaleheight})) we then find that the
line-of-sight crosses the disk at a height $H$ of about 5.3 AU. The
inclination $i$ is then found by solving
\begin{equation}
\tan(90\degr -i) = \frac{H}{R_0 - D/2}
\end{equation}
yielding an inclination of 60\degr.
This shows that the variation in the optical depth can be understood
in terms of the scale height of a hydrostatic disk.  It also shows
that the disk must be truncated to about 2-3 pressure scale heights,
consistent with what we found in section \ref{sec:grain-drift}.
corresponding to an $H/R$ of 0.58.  However, the dynamic truncation of
the disk means that the density distribution in not a perfect
Gaussian.  The inclination derived above is therefore an
approximation. \\

The Wall model we presented in this paper is therefore not only capable
of explaining the blackbody shape of the IR continuum, but can also
explain in a consistent way the variations in the UV absorption tied
to the orbital variations. In that case, the chemical composition of
the dust grains and the exact geometric configuration of the different
grain species observed through either the UV absorption or the
features in the ISO-SWS spectrum remains a matter of debate. The total
mass in the circumbinary disk which is required to make the optical
depths sufficiently high (see Sect.~\ref{sec:radial-tau}) on the other
hand might well be compatible with the total mass lost during the
evolution of HR 4049 on the RGB and/or the AGB taking into account
that no mass has left the system. \\
The main difficulty with the Wall model is accurately reproducing the
exact blackbody shape which we unambiguously observe. As was shown in
Sect.~\ref{sec:diffusion}, leakage through the top, bottom and rear
surfaces of the disk still produces a significant contribution to the
flux at long wavelengths, even for very high optical depths.

\subsection{The metallicity of HR4049}
We have also derived properties concerning the disk stability.  We
showed that the central part of a massive disk can withstand the
radiation pressure acting on the dust grains, because the radiation
force will only act on the innermost grains, while the rest of the
disk does not receive direct stellar radiation.  It is obvious, that a
much more thorough study of the dynamics of such a disk would be very
interesting.  Such a study requires two-fluid hydrodynamics and
radiative transfer involving large optical depth.  We may however
speculate, that a study of the true dynamics of such a disk may
provide a natural explanation for the extremely low metallicity of
HR4049.  Envision the disk filled with small grains.  The radiation
pressure of the star will push the grains into the inner rim of the
disk, leading to very high dust densities there.  The dust grains will
collide with the gas and provide an outward force to the entire
medium.  But since the outer disk does not receive direct stellar
light, it is not moving radially.  Therefore, either a large radial
pressure gradient or substantial sub-Keplerian rotation must balance
the radiation force.  However, we have shown in section
\ref{sec:grain-drift} that the grains will slowly drift outwards
through the gas, reducing the dust content in the innermost layers.
Once a layer of gas has become dust free, the radiation force acting
on this layer will become very small.  The layer falls inwards, driven
either by the gas pressure gradient, or by its sub-Keplerian rotation.
The falling gas will be free of dust and therefore have very low metal
abundances.  If this gas can eventually be reaccreted by the primary
of HR4049, it will form a metal-poor photospheric layer on the star.
Since the disk may contain a significant fraction of a solar mass,
a substantial amount of metal free gas is available for this
mechanism.

\subsection{Fat circumstellar disks in post-AGB binary stars}

Geometrically thick circumstellar disks are not uncommon in post-AGB
stars.  Similar to HR\,4049 are objects with a detected binary with a
rather small orbital period implying strong binary interaction during
evolution.  A famous example is the Red Rectangle (HD 44179), where
the disk is even resolved \citep{1995ApJ...443..249R}.  The Red
Rectangle also shows variability in phase with the orbital motion of
the binary, but the variability is entirely grey.
\citet{1996A&A...314L..17W} explain the variability  in this source
with a variable angle under which the stellar light is scattered
towards the observer.  This model also requires the presence of a fat
disk around the system, similar to our proposed wall model.  Another
example is HD\,213985, which shows the same photometric behavior as
HR\,4049 \citep{1995Ap&SS.224..357W}.  The link to the central binary
characteristics is less clear for objects where the disc is resolved
like the Egg Nebula \citep{1998ApJ...493..301S}.  The nature of the
central binary in this object is not yet known.  

However, none of the above sources shows the perfect blackbody shape
of the excess emission.

\section{Conclusions}
The Spectral Energy Distribution of HR 4049 shows three significant
keys which can be used to determine the basic properties of its
circumstellar dust: A variable, non-grey circumstellar extinction, an
infrared excess which amounts to 1/3 of the stellar flux and the
perfect blackbody shape of the excess between 1\micron and 850\micron.
We conclude that the bulk of the circumstellar matter must be in a
disk with large scaleheight $H/R\sim 1/3$.  We have considered large
grains in a marginally optically thick disk as a possible explanation
and find that such a structure is subject to very tight constraints on
the radial extent and optical depth of the disk, and that the large
scale height is difficult to support for an extended time.  We
propose, as an alternative solution, the wall model which is a fat disk
with a radial optical depth in excess of 10$^{4}$.  The wall model
fulfills many of the requirements, with the only significant
limitation that the optical depth must be very high to suppress excess
flux at 850\micron.  The wall can explain the observed variable
extinction in a very consistent way, while the big-grain model
requires a separate component for this observation.  In the context of
binary evolution, the formation of a massive circum-binary disk can be
qualitatively understood as the result of tidal interaction during the
AGB phase of the primary.  The wall model is therefore an attractive
candidate to describe the CS environment of HR\,4049.

In the near future, the MIDI interferometer at the VLT should be able
to resolve the disk around HR\,4049.  Wavelength dependent
interferometric measurements will give information about the location
of PAH emission and continuum, and about the location of the inner
disk rim.  HR\,4049 has fascinated astronomers now for about 20 years
- there is more to come.

\begin{acknowledgement}
  We would like  to thank Rens Waters, Xander Tielens and Jorge
  Jimenez Vicente for stimulating and insightful discussions.  We
  would like to thank Remo Tilanus for the SCUBA observations
  and the data reduction. CD and
  JC acknowledge the financial support from NWO Pionier grant
  6000-78-333.  CPD and CD acknowledge support from the European
  Commission under TMR grant ERBFMRX-CT98-0195 (`Accretion onto black
  holes, compact objects and protostars').
\end{acknowledgement}



\end{document}